\documentclass[preprint,showpacs,preprintnumbers,amsmath,amssymb]{revtex4}

\usepackage{amsmath}
\usepackage{mathrsfs}
\usepackage{graphicx}
\usepackage{dcolumn}
\usepackage{bm}
\usepackage[colorlinks, dvipdfm, citecolor=blue]{hyperref}
\usepackage{booktabs}
\begin{document}

\title{ An {\it ab initio} design of cluster-assembled silicon nanotubes}

 \author{Lingju Guo}
 \author{Xiaohong Zheng}
 \author{Chunsheng Liu}
 \author{Zhi Zeng}%
\altaffiliation{Corresponding author. E-mail: zzeng@theory.issp.ac.cn}
\affiliation{%
Key Laboratory of Materials Physics, Institute of Solid State
Physics, Chinese Academy of Sciences, Hefei 230031, P.R. China
}%

\begin{abstract}
Density functional calculations were performed to systematically
study a series of finite and infinite cluster-assembled silicon
nanotubes (SiNTs). One-dimensional SiNTs can be prepared by proper
assembly of hydrogenated cage-like silicon clusters to form
semiconductors with a large band gap, and their electronic
properties can be accurately tuned by transition metal doping in the
center of the tubes. Specifically, doping with Fe made the SiNTs
metallic and magnetic materials. More interestingly, a metal to
half-metal transition was observed with increasing tube radius in
Fe-doped SiNTs, which demonstrates that SiNTs doped with magnetic
elements may find important applications in spintronics.


\end{abstract}

\maketitle

\section{INTRODUCTION}
Since the discovery and application of carbon fullerenes and carbon
nanotubes(CNTs) \cite{fullerene}, stable cage and tube-like
structures have attracted a great deal of attention. These
structures have generated great interest in creating analogous
structures from other elements that are suitable for applications in
nanodevices. Silicon and carbon are members of the same group in the
periodic table, suggesting a potential ability to form similar
structures. Furthermore, due to the fundamental importance of
silicon in present-day integrated circuits, substantial efforts have
focused on investigating nano-scale forms of silicon, both for the
purpose of further miniaturizing the current microelectronic devices
and in the hopes of unveiling new properties that often arise at the
nano-scale level\cite{JPCM_16_1373}. However, it is difficult to
form cages or tubes like carbon fullerenes or nanotubes purely with
Si atoms because silicon does not favor the $sp$$^{2}$ hybridization
that carbon does. Carbon normally forms strong $\pi$ bonds through
$sp^{2}$ hybridization, which can facilitate the formation of
two-dimensional spherical cages (or planar structures such as
benzene and graphene). Silicon, on the other hand, usually forms
covalent $\sigma$ bonds through $sp^{3}$ hybridization, which favor
a three dimensional diamond-like structure.

Fortunately, Si cage clusters can be synthesized by adding suitable
foreign atoms to terminate the dangling Si bonds that inherently
arise in cage-like networks. Many
researchers\cite{jctn257,cms1,prb195417} reported that transition
metal(TM) atoms are the most suitable element for cage formation due
to their $d$ band features. In addition, rare earth atoms have been
doped into silicon
cages\cite{jcp084711,epjd343,prb125411,prb115429}. Another way to
stabilize the Si cage is to use hydrogens to terminate the cluster
surface\cite{prb075402,prb155425}, which is similar to the
dodecahedral C$_{20}$H$_{20}$.

Several hollow and nonhollow silicon nanotube structures have been
proposed and theoretically characterized in recent years
\cite{pnas2664, jmst127, prl265502, nl301, nl1243, nano109, jmc555,
njp78, prl146802, cpl81, prb075420, prb195426, jpcb7577, jpcb8605,
prb9994, jpcc5598, prb11593, prb205315, prb075328, ss257, prb193409,
jpcc16840, prl1958, pssr7, prb155432, prl792, jpcc1234}. These
structures, which were proposed based on intuition or the behavior
of similar materials, include the following:

(1) Regular polyhedron stacking nanotubes \cite{pnas2664},

(2) Surface capped polygonal prism stacking nanotubes
\cite{jmst127},

(3) Polycrystalline forms of nanowires \cite{prl265502},

(4) Metal encapsulated polygonal prism nanotubes
\cite{nl301,nl1243,nano109,jmc555,njp78,prl146802},

(5) Carbon nanotube like structures
\cite{cpl81,prb075420,prb195426,jpcb7577,jpcb8605,prb9994,jpcc5598},

(6) Fullerene-based structures \cite{prb11593,prb205315},

(7) Metal centered fullerene-based structures \cite{prb075328},

(8) Hydrogenated single-wall silicon CNT like nanotubes
\cite{ss257,prb193409,jpcc16840},

(9) Exohydrogenated fullerene-like structures \cite{prl1958,pssr7},

(10) Multiwall nanotubes \cite{prb155432,prl792,jpcc1234}.

In the cases of (1), (3), (5) and  (6), similar to pure silicon
cages, it was reported to be difficult to form hollow single wall Si
nanotubes because of the $sp^{3}$ nature of Si atom. Cases
(4)~and~(7) proposed that metal doping may be a good way to
terminate the dangling bond of silicon bond in the tube, but this
approach was limited to tubes with very small radii($R$$\leq$1nm).

However,  if the lateral surfaces of the dangling bonds are
terminated by hydrogens, the resulting cage-like silicon clusters
may be perfect building blocks for Si nanotubes. Theoretically,
every Si atom has three neighboring Si atoms in the nanotube, and
these have an $sp$$^{3}$ type bond nature with one H terminating the
dangling bond. Moreover, the success of this approach is suggested
by the fact that in experiments, the surfaces of Silicon
nanowires(SiNWs) and silicon nanotubes(SiNTs) are always passivated
by hydrogen atoms \cite{science1874,jpcb8605} or by silicon oxide
layers \cite{am1219,am1172,am564,prl116102}.

This work presents a theoretical study of hydrogenated cluster
assembled silicon nanotubes using density functional theory(DFT)
calculation. A series of finite and infinite silicon nanotubes
assembled by hydrogenated cage-like clusters was obtained.
Additionally, the electronic and magnetic properties of SiNTs can be
accurately tuned by doping impurities at the center of the hollow
tube. One interesting result was that a metal to half-metal
transition was observed in Fe-doped SiNTs  with the increasing of
tube radius.

\section{COMPUTATIONAL DETAILS AND MODEL DESIGN}\label{cd}

All theoretical computations were performed with the DFT approach
implemented in the Dmol$^{3}$ package \cite{jcp92,jcp113}, using all
electron treatment and the double numerical basis including the
$d$-polarization function(DNP)\cite{jcp92}. The exchange-correlation
interaction was treated within the generalized gradient
approximation(GGA) using BLYP functional. Self-consistent field
calculations were performed with a convergence criterion of
2$\times$10$^{-5}$ Hartree on total energy.

The $Si_{n}H_{n}$~($n$=16, 20, 24, 28, 32) clusters were optimized
first, and some of the initial structures in our work were based on
the results reported in Refs.~\onlinecite{prb075402} \&
\onlinecite{prb155425}. After obtaining stable single clusters, we
took these clusters as basic units (keep them as original) and
stacked the clusters with a certain $n$ along the axis of symmetry
to construct finite nanotubes.

For infinite one dimensional nanotubes, the periodically repeated
units were placed in supercells. The supercell is cubic in geometry
with the dimensions 25{\AA}$\times$25{\AA}$\times$$L$$_{z}$, where
 the direction of $L$$_{z}$  was defined as the axial direction of the nanotubes.
Periodic boundary conditions were employed along the nanowire axis
to create, in effect, continuous wires. Meanwhile, a sufficiently
large vacuum region was introduced along the other directions is
configured between the wires.

\section{RESULTS AND DISCUSSIONS}

\subsection{Structures of $Si_{n}H_{n}$ clusters and finite nanotubes }\label{f}

The fully optimized structures of $Si_nH_n$ ($n$=16, 20, 24, 28, 32)
clusters that have been fully optimized are shown in Fig.
\ref{fig1}. It was very interesting to see that all these structures
shared the following common characteristics: 1. All of them were
fullerene-like hollow structures. 2. Each Si atom had three Si
neighbors, with one H atom saturating the dangling bond of each
silicon atom outside the cage to fulfill an $sp^3$ type
hybridization bond.  3. All structures consisted of $\frac{n}{2}$
pentagons and two other polyhedrons at the two ends, with the edge
number of these two polyhedrons as $\frac{n}{4}$. Meanwhile, these
two polyhedrons were parallel to each other and there was a relative
angle of $\frac{4\pi}{n}$ between them. Thus each vertex atom of one
polyhedron fell exactly in the middle of one edge in the other
polyhedron. Specifically, for $Si_{20}H_{20}$, the cage was composed
of 12 pentagons, which was very similar to the structure of carbon
fullerene $C_{20}H_{20}$.
 Structures of $Si_{16}H_{16}$, $Si_{24}H_{24}$ and $Si_{28}H_{28}$
have been widely discussed in very recent years
\cite{prb075402,prb155425} and the structural information we
obtained was consistent with these reports.

Taking these original clusters original as basic units, we stacked
them along the central axis of the cage to form finite nanotubes.
The two adjacent cages shared the same polyhedron. We noted that for
the shared polyhedron, there was no need for have hydrogen
saturation because each Si atom already had 4 Si neighbors and thus
the $sp^3$ hybridization bond type was fulfilled. The molecular
formula of the finite tube was classified as
$Si_{m(3k+1)}H_{2m(k+1)}$, where the measure of radius $m$ was the
number of atoms of one shared polyhedron and the measure of length
$k$ was the number of repeated units. After full optimization, for
one certain value $m$($m$=4, 5, 6, 7), and for $k$ range from 2 to 4
concerned in the present work, the tube was always straight and
stable. Furthermore, if the number of repeated units $k$ was fixed,
the length of the tubes decreased with the increasing $m$. The
angles of H-Si-Si and Si-Si-Si inside the repeated units were all
about 109$^{\circ}$, which were very close to the 109.5$^{\circ}$ of
$sp^3$ , but the Si-Si-Si angle~(angle $\alpha$ in Fig.~\ref{fig2}
(b)) between two units became smaller and smaller with the
increasing of tube radii (from 127.8$^{\circ}$ of $m$=4 to
99.0$^{\circ}$ of $m$=8).

Even though, the growth direction and the achievable length of the
nanotubes are the main experimental concerns, previous theoretical
studies have paid little attention to these issues. Here, we chose
$Si_{5(3k+1)}H_{10(k+1)}$ nanotubes to examine if straight tubes
were more stable than bent nanotubes. As shown in Fig.~\ref{fig2},
 $Si_{50}H_{40}$ ($m$=5, $k$=3)(Fig.~\ref{fig2}~(a)) had
 an isomer (a$_{1}$) (Fig.~\ref{fig2}~($a_{1}$)) that was bent to
 120$^{\circ}$ from 180$^{\circ}$, but the total energy (E$_{T}$) of ($a_{1}$) was 0.02eV higher than
 that of the straight tube ($a$).
 Likewise,
 $Si_{65}H_{50}$ ($m$=5, $k$=4) has two isomers $b_{1}$(Fig.~\ref{fig2}~($b_{1}$))
  and $b_{2}$ (Fig.~\ref{fig2}($b_{2}$)),
 where the structure $b_{1}$ was a tube in which the unit at the end was bent,
 and structure $b_{2}$ was distorted further. These isomers were less stable
 because the total energy of ($E_{T}$) was higher than that of the linear
 structure  by 0.021$eV$ or 0.034$eV$,
respectively. For hydrogenate silicon tubes, if one tube was bent by
an angle, the distance between H atoms of adjacent units would
decrease and the Coulomb repulsion between them would makes the
total energy greater than the straight energy. Consequently, the
linear structure was relatively more stable than bent structures.

\subsection{ELECTRONIC STRUCTURES}

In order to measure the relative stability of the tubes as well as
the influence of the length and width, we calculated the averaged
binding energy~($E_{b}$) and dissociation energies~($DE$) of these
tubes. The $E_{b}$ and DE for finite
 tubes were defined by the following formulae:
\begin{equation}
\begin{split}
E_{b}({k})=\{m(3k+1)E_{T}&(Si)+2m(k+1)E_{T}(H)-E_{T}[Si_{3k+1}H_{2m(k+1)}]\}/{(5mk+3m)}\\
&(m=4-8;  ~k=1-4) \label{a}
\end{split}
\end{equation}

\begin{equation}
\begin{split}
\ DE({k})&=E_{T}(SiH)-E_{T}[Si_{3k+1}H_{2m(k+1)}]\\
&(m=4-8; ~k=1-4~)
\end{split}\label{a}
\end{equation}
where $E_{T}(Si)$, $E_{T}(H)$, $E_{T}(SiH)$ and
$E_{T}$[$Si_{m(3k+1)}H_{2m(k+1)}$] represent the total energies of
the Si atom, H atom, SiH dimer and the $Si_{m(3k+1)}H_{2m(k+1)}$
tube, respectively.

As illustrated in Fig.~\ref{fig3}~(a), the binding energy $E_{b}(k)$
of the finite tube increases gradually as the length increased for a
certain $m$. This correlation indicates that the tube became
increasingly stable as it became longer. On the other hand, the
stability of the tubes did not depend linearly on the tube's radius,
which was represented by $m$. For any $k$, the most stable tubes
were at $m$=5 or $m$=6.

DE is the energy needed to remove a Si-H dimer from the end of a
tube,  and this parameter provided another method to probe the tube
stability. The curves of DE~(Fig.~\ref{fig3}~(b)) indicate that it
was more difficult to remove a Si-H dimer from the tubes with $m$=5
and 6 because these two specific tubes satisfied $sp^{3}$
hybridization well. This is another indication of  the fact that
tubes with $m$=5 or 6 are the most stable.

 In addition, Fig.~\ref{fig3}~(c) plots the variation of the
 highest occupied molecular orbital~(HOMO) and lowest unoccupied molecular orbital~(LUMO)
  gaps of the finite tubes. The quantum confinement
 concept demands a larger gap for a smaller size. The authors in Ref.~\onlinecite{prb075402} ever argued
 that the quantum confinement concept was not fully applicable to saturated systems
 such as $Si_{n}H_{n}$ cages. Our work also provides evidence on the
 limitation of the quantum confinement concept in interpreting the
 gaps of hydrogenated silicon nanotubes.

As seen in Fig.~\ref{fig3}~(c), when $m$ is fixed, the gaps between
 HOMO-LUMO became smaller with increasing length. However, there was a big
 jump in the gap from a single cluster to a two-unit tube.
 This difference could come from a hybrid bond around the Si$_{m}$ cylinder
joints which was different from the fully hydrogenated silicon atoms
in a single cluster. Moreover, the Coulomb repulsion between
hydrogen atoms of an adjacent unit could play a role in defining
gaps, as explained below.

The tubes of $Si_{4(3k+1)}H_{2\times4(k+1)}$,
 $Si_{5(3k+1)}H_{2\times5(k+1)}$ and $Si_{6(3k+1)}H_{2\times6(k+1)}$
 had
 similar HOMO-LUMO gaps due to analogous structural parameters including bond angle, length and
 width. The gaps of larger radii tubes $Si_{7(3k+1)}H_{2\times7(k+1)}$ and
$Si_{8(3k+1)}H_{2\times8(k+1)}$ tubes were much smaller than those
of the smaller radii tubes $Si_{4(3k+1)}H_{2\times4(k+1)}$,
$Si_{5(3k+1)}H_{2\times5(k+1)}$ and $Si_{6(3k+1)}H_{2\times6(k+1)}$.
Similar to Ref.~\onlinecite{prb075402}, we estimated the effective
volume per electron $V_{eff}^{(e)}$ by drawing two concentric
cylinders with $R_{H}$ and $R_{Si}$ in Fig.~\ref{fig2}~(b) up to the
layer of hydrogen and silicon.

The $V_{eff}^{(e)}$ was defined as:
\begin{equation}
 V_{eff}^{(e)}=\frac{1}{N_{e}}\times \pi \times L~[(\frac{1}{2}R_{H})^2-(\frac{1}{2}R_{Si})^2] \label{c}
\end{equation}
where $N_{e}$ was the number of electron and $L$ was the length of
the cylinder (Fig.~ \ref{fig2}~(b)).

$V_{eff}^{(e)}$ is the measures of electron "confinement", which
would lead to the same order of gap performance according to the
quantum confinement concept.  Hence, in our work, the order of
effective volume per electron of $V_{eff}^{(e)}$[$m$=7] $<$
$V_{eff}^{(e)}$[$m$=6] $<$ $V_{eff}^{(e)}$[$m$=8] $<$
$V_{eff}^{(e)}$[$m$=4] $<$ $V_{eff}^{(e)}$[$m$=5], as in
Table~\ref{table1}, would have resulted in a HOMO-LUMO gap order of
$\Delta_{G}$~[$m$=7]$>$ $\Delta_{G}$~[$m$=6] $>$
$\Delta_{G}$~[$m$=8] $>$ $\Delta_{G}$~[$m$=4] $>$
$\Delta_{G}$~[$m$=5]. However, as shown in Fig.~\ref{fig3}(c) and
Table~\ref{table1}, the HOMO-LUMO gap order was $\Delta_{G}$~[$m$=6]
$>$ $\Delta_{G}$~[$m$=4] $>$ $\Delta_{G}$~[$m$=5] $>$
$\Delta_{G}$~[$m$=7] $>$ $\Delta_{G}$~[$m$=8]. The HOMO-LUMO gap
order only followed the rule of quantum confinement for small radii
tubes($m$=4, 5, 6). The limitation of quantum confinement for large
radii tubes ($m$=7, 8) may arise from the fact that these tubes have
a smaller distance between the H-H bonds of neighboring
units($D_{H}$ in Fig.~\ref{fig2}~(b)). For example, $D_{H}$ of
4.206{\AA} for $m$=4 compared to that of 2.723{\AA} or 2.567{\AA}
for $m$=7 or 8. That is to say, for one-dimensional finite
hydrogenated silicon nanotubes, quantum confinement only worked for
small radii tubes.

To summarize, in addition to the quantum confinement, the effects of
suitable bond angle($\angle\alpha$ in Fig.~\ref{fig2}~(b)), and
Coulomb repulsion between H atoms also played important roles in
affecting the HOMO-LUMO gaps of $Si_{m(3k+1)}H_{2m(k+1)}$ systems.

\subsection{INFINITE TUBES}
The increased stability of finite nanotubes with an increasing
number of units allowed us to examine further the stability of the
infinite nanotubes specifical with a stoichiometric $Si_{6m}H_{4m}$.
This infinite tube was built from finite $Si_{m(3k+1)}H_{2m(k+1)}$
by removing a $Si_{m}$$H_{m}$ on one side and $m$ H atoms on the
other side to form a repeated unit. Fig.~\ref{fig4} shows two
repeated cells of the infinite tubes with different radii. All of
these tubes had the stacking of SiH cages in common, with the wire
axis lying in the center. The axis passed through the centers of
buckled $Si_{m}$ rings, and two adjacent cages shared one ring.

Full structure relaxation indicated that the infinite nanotubes had
similar geometric structures to finite ones, but the tube lengths
had slightly changed. The lenths of $Si_{6\times4}H_{4\times4}$ and
$Si_{6\times5}H_{4\times5}$ tubes became 0.08{\AA} and 0.11{\AA}
longer than the finite ones, but for $Si_{6\times6}H_{4\times6}$,
$Si_{6\times7}H_{4\times7}$ and $Si_{6\times7}H_{4\times7}$ the
lengths became 0.075{\AA}, 0.212{\AA} and 0.365{\AA} shorter. The
tube widths and Si-H  bond lengths were almost the same as for the
finite ones.

 The binding energies of the infinite tube shown in Table~\ref{table2} were slightly larger than
those of finite tubes, which meant that it was possible to
synthesize  tubes long enough. The silicon tube had a relatively
large band gap($\Delta_{g}$)~(2.296eV-3.009eV), which implied that
it was a wide gap semiconductor. The band gap of the tube was
inversely proportional to the radius.

The band gaps of carbon nanotubes always decrease in an oscillatory
behavior with increasing radius because of the $\pi^{\ast}$ and
$\sigma$ hybridization with a small radius and a large curvature.
However, for exhydrogenated single-wall carbon nanotubes(SWCNT), the
gap decreases monotonously \cite{prb075404}. This difference can be
easily understood  by the fact that silicon atoms have $sp^{3}$-type
bond properties. The band-gap trend of silicon nanotubes  is similar
to that of hydrogenated SWCNTs as described in other
reports~\cite{prb075328, prb193409, prb155435} on Si nanotubes or
nanowires.

\subsection{IRON DOPED INFINITE TUBES}
Endohedral doping is an effective way to tune the properties of cage
or tube-like clusters. Therefore, we next investigated the effect of
endohedral doping on the geometric stability, conductivity and
magnetism of the silicon nanotubes. Specially, Fe doping was
systematically analyzed. The stability was robust when  Fe atoms
were inserted at the center of the hollow tube(Fig.~\ref{fig4}).
However, the lattice underwent a very small expansion and exiting
large band gap in silicon nanotube disappeared. Subsequently, the
doped tube turned into a metal.

In previous reports, many types of metal atoms were inserted at the
center of the silicon hollow nano-clusters \cite{cms1,prb195417} or
nanotubes \cite{nl1243, nl301, prb075328, jmc555}, but the primary
role of the doping atom was to saturate the dangling bond of silicon
cages or tubes. Therefore, the central atom had a strong
hybridization with the exterior silicon cage, and the magnetic
moments always became very small, or at times completely quenched.
However, in the case presented here, the outside cage was saturated
by hydrogen atoms. Thus, the Fe atom had only a weak interaction
with the silicon atoms. As seen in Table~\ref{table2}, the doped
tube always maintained a relatively large magnetic moment that
originated from the $d$-electron of the Fe atom, and the variation
of the Fe moment increased as the radius increased.

The most intriguing phenomenon occurred when the radii of the
Fe-doped hydrogenated silicon tubes increased: a metal to half-metal
transition was observed from the spin resolved density of states
(DOS) for the $\alpha$ and $\beta$ channels of nanotubes,
 as shown in Fig.~\ref{fig5}. From $m$=4 to $m$=6
(Fig.~\ref{fig5}(a)--(c)), both the $\alpha$ spin and the $\beta$
spin contributed to the DOS at the Fermi level. With the increase in
$m$, the DOS peak of the $\alpha$ spin at Fermi level shifted down,
while the peak of the $\beta$ spin below the Fermi level shifted up
towards the Fermi level. When it comes to $m$=7 (Fig.~\ref{fig5}
(d))and $m$=8 (Fig.~\ref{fig5} (e)), the $\alpha$ spin states
disappeared at the Fermi level, and a band gap appeared for this
spin channel. For the $\beta$ spin channel, a DOS peak still emerged
at the Fermi level. In other words, the nanotube turned into
 an insulator for $\alpha$ spin and a conductor for the $\beta$ spin.
From the spin-resolved DOS contributed by Fe, Si and H atoms shown
in Fig.~\ref{fig5}(f), we found that the centered Fe atoms mainly
contributed to the total number of $\beta$ spin states, and Si atoms
made only very modest contributions to the states at the Fermi
level.

The above phenomenon resulted from the hybridization between Fe and
silicon atoms. For small radii of $m$=4, Fe atoms have a  stronger
hybridization with the outside silicon atoms, but the hybridization
becomes smaller as the radius increases. When $m$=7 and 8, the
coupling between the Fe atom and the tube becomes so weak that the
Fe atom is almost an isolated atom. Thus, the atomic properties of
Fe atom were recovered to a large extent. The electron configuration
of Fe is 3$d$$^{6}$4$s$$^{2}$. According to Pauli exclusion and
Hund's rule, five 3$d$ electrons would occupy five $d$ orbitals with
$\alpha$ spin, while the last 3$d$ electron will occupy $d$ orbitals
with $\beta$ spin. When such atoms are weakly coupled together to
form a linear chain, highly spin-polarized behavior can result.

Such behavior was also been predicted in Durgun's
work\cite{prl256806} which found that hydrogen passivated silicon
nanowires could exhibit the half-metallic state when decorated with
specific transition metal(TM) atoms. This situation was very similar
to the half-metallicity introduced by the weakly coupled carbon
atomic chain in single-walled carbon nanotubes\cite{apl163105}. In
this case, the carbon atoms in the chain have a very weak
interactions with their neighbors and the atomic behavior is largely
recovered.

The large magnetic moments of the Fe doped silicon nanotubes and
their half-metallic behavior are promising for application of these
materials in magnetic devices and spintronic applications.

\section{CONCLUSIONS}

A series of finite and infinite hydrogenated silicon nanotubes were
systematically studied by performing first-principles DFT-GGA
calculations. Our results demonstrated that one-dimensional SiNTs
$Si_{m(3k+1)}H_{2m(k+1)}$ can be built by stacking $Si_{k}H_{k}$
clusters along the central axis of the cage. The finite tubes had
large HOMO-LUMO gaps. When $m$=5 and 6, the tubes were more stable
than other sizes because these tubes had the most suitable bond
angles.

Infinite tubes had similar geometric structures to the finite tubes
and presented wide gap semiconductivity due to their large band
gaps.
 The band gap decreased monotonously as the radius increased.
Endohedral Fe doping greatly modified the properties of the silicon
nanotubes. For example, Fe doping can tune the silicon nanotube from
semiconductor to metal. Additionally, a metal to half-metal
transition was observed in Fe-doped SiNTs with an increasing tube
radius. These silicon nanotubes could be useful for nanoelectronic
devices. In particular, a half-metallicity of the Fe-doped silicon
nanotubes may find important application in spintronics.

\section{ACKNOWLEDGMENTS}

This work was supported by the National Science Foundation of China
under Grants No. 10774148 and No. 10904148, the special Funds for
Major State Basic Research Project of China(973) under grant No.
2007CB925004, 863 Project, Knowledge Innovation Program of the
Chinese Academy of Sciences, and Director Grants of CASHIPS. Some of
the calculations were performed at the Center for Computational
Science of CASHIPS and at the Shanghai Supercomputer Center.




\begin{small}

\section{Figure and Table captions}
\underline{Table \ref{table1}}. HOMO-LUMO gap($\Delta_{G}$) and
effective volume per electron $V_{eff}^{(e)}$ of
 $Si_{m(3\times4+1)}H_{2m(4+1)}$ ($m$=4, 5, 6, 7, 8) tubes.

 \underline{Table \ref{table2}}. Binding
energy~($E_{b}$) of infinite SiH tubes and Fe-doped SiH tubes, the
band gap ($\Delta_{g}$)of SiH tubes, and the spin moments($M$) per
Fe atom and of Fe-doped SiH tubes.

\underline{$Fig$. \ref{fig1}}. (Color online) Top view of the
optimized $Si_{n}H_{n}$ clusters (the 1st row) and side view of
cluster-assembled nanotubes $Si_{m(3k+1)}H_{2m(k+1)}$ (all the
rest).

\underline{$Fig$. \ref{fig2}}. (Color online) Structures of
$Si_{5(3k+1)}H_{3\times5(k+1)}$ ($k$=3, 4) tubes.

\underline{$Fig$. \ref{fig3}}. (Color online) The averaged binding
energies, dissociation energies and HOMO-LUMO gaps of finite
$Si_{m(3k+1)}H_{2m(k+1)}$ tubes.

\underline{$Fig$. \ref{fig4}}. (Color online) Side views of two unit
cells of $Si_{6m}H_{4m}$ nanotubes, (a) pristine SiH nanotubes, (b)
Fe-doped SiH nanotubes, where the dark atoms in the center of the
tube represent Fe atoms.

\underline{$Fig$. \ref{fig5}}. (Color online) $\alpha$ and $\beta$
spin resolved DOS of Fe-doped SiH nanotubes($Si_{6m}H_{4m}$).
(a)--(e): total $\alpha$ and $\beta$ spin DOS of
$Si_{6\times4}H_{4\times4}$, $Si_{6\times5}H_{4\times5}$,
$Si_{6\times6}H_{4\times6}$, $Si_{6\times7}H_{4\times7}$, and
$Si_{6\times8}H_{4\times8}$, (f): partial $\alpha$ and $\beta$ spin
DOS of $Si_{6\times7}H_{4\times7}$.

\newpage

\begin{table}
\caption{\label{table1} \LARGE{Guo $et$ $al$.}}
\begin{ruledtabular}
\begin{tabular}{ccc}
$Si_{m(3k+1)}H_{2m(k+1)}$&$\Delta_{G}$~(eV)&$V_{eff}^{(e)}$~({\AA}$^{3}$/electron)\\
\hline
$Si_{52}H_{40}$ ($m$=4, $k$=4) & 2.934 & 6.748\\
$Si_{65}H_{50}$ ($m$=5, $k$=4) & 2.882 & 6.760\\
$Si_{78}H_{60}$ ($m$=6, $k$=4) & 3.006 & 6.653\\
$Si_{91}H_{70}$ ($m$=7, $k$=4) & 2.592 & 6.632\\
$Si_{104}H_{80}$ ($m$=8, $k$=4)& 2.257 & 6.681\\
\end{tabular}
\end{ruledtabular}
\end{table}

\begin{table}
\caption{\label{table2} \LARGE{Guo $et$ $al$.}}
\begin{ruledtabular}
\begin{tabular}{cccccccc}
&&&&\multicolumn{2}{c}{$M$~($\mu_{B}$)}\\
\cline{5-6}
m&$E_{b}$(SiH)~(eV)&$E_{b}$(FeSiH)~(eV)&$\Delta_{g}$~(eV)&per Fe&total\\
\hline
4&3.476&3.348&3.009&2.822&6.900\\
5&3.482&3.411&2.765&3.183&8.000\\
6&3.500&3.424&2.921&3.395&8.000\\
7&3.479&3.415&2.557&3.538&8.000\\
8&3.448&3.387&2.296&3.722&7.967\\
\end{tabular}
\end{ruledtabular}
\end{table}

\newpage

\newpage
\begin{figure}
\centering
\includegraphics[bb =40 0 900 790,width=1.60\textwidth]{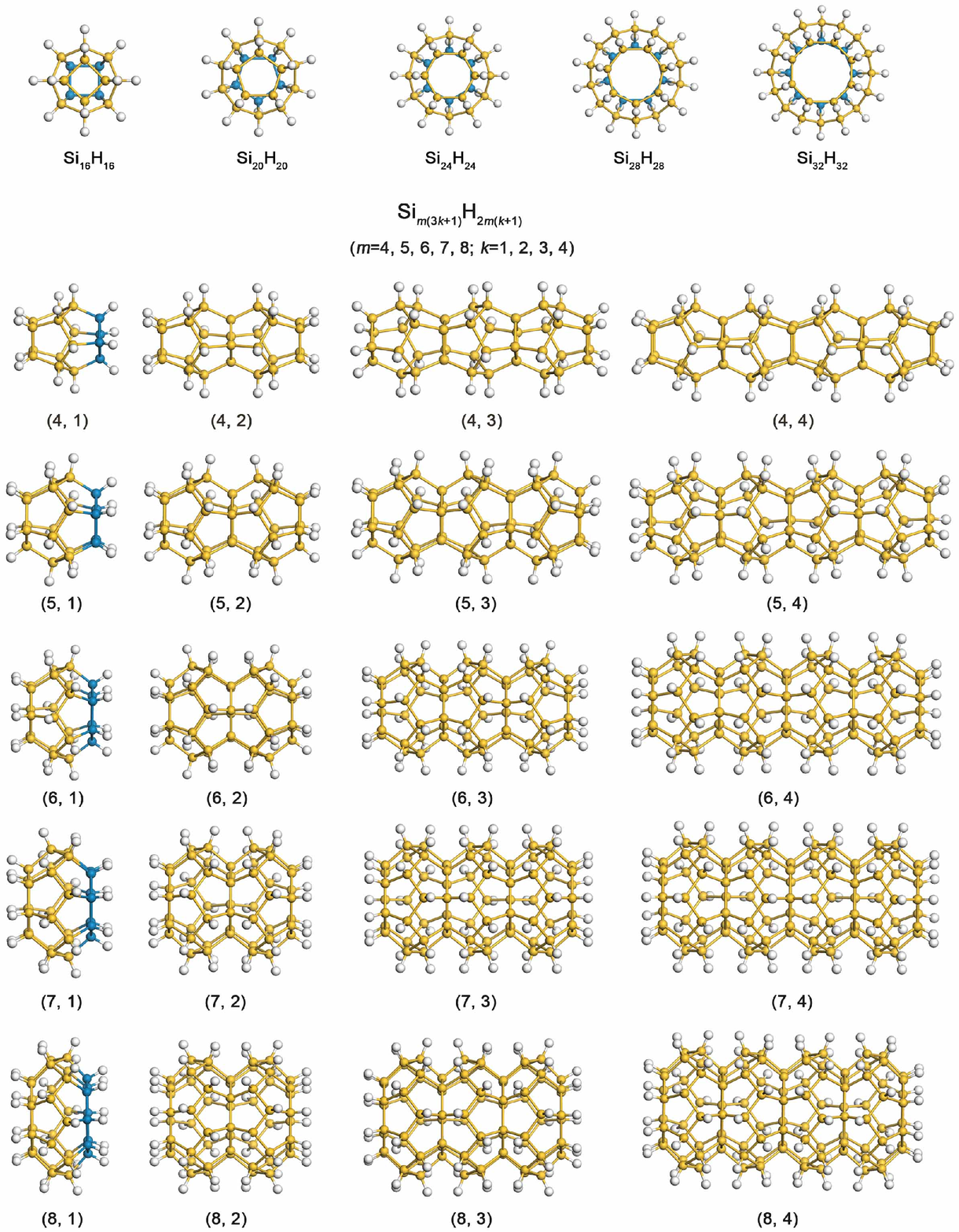}
\caption{\label{fig1}}
\end{figure}

\newpage
\begin{figure}
 \centering
\includegraphics[bb =50 0 650 500,width=1.50\textwidth]{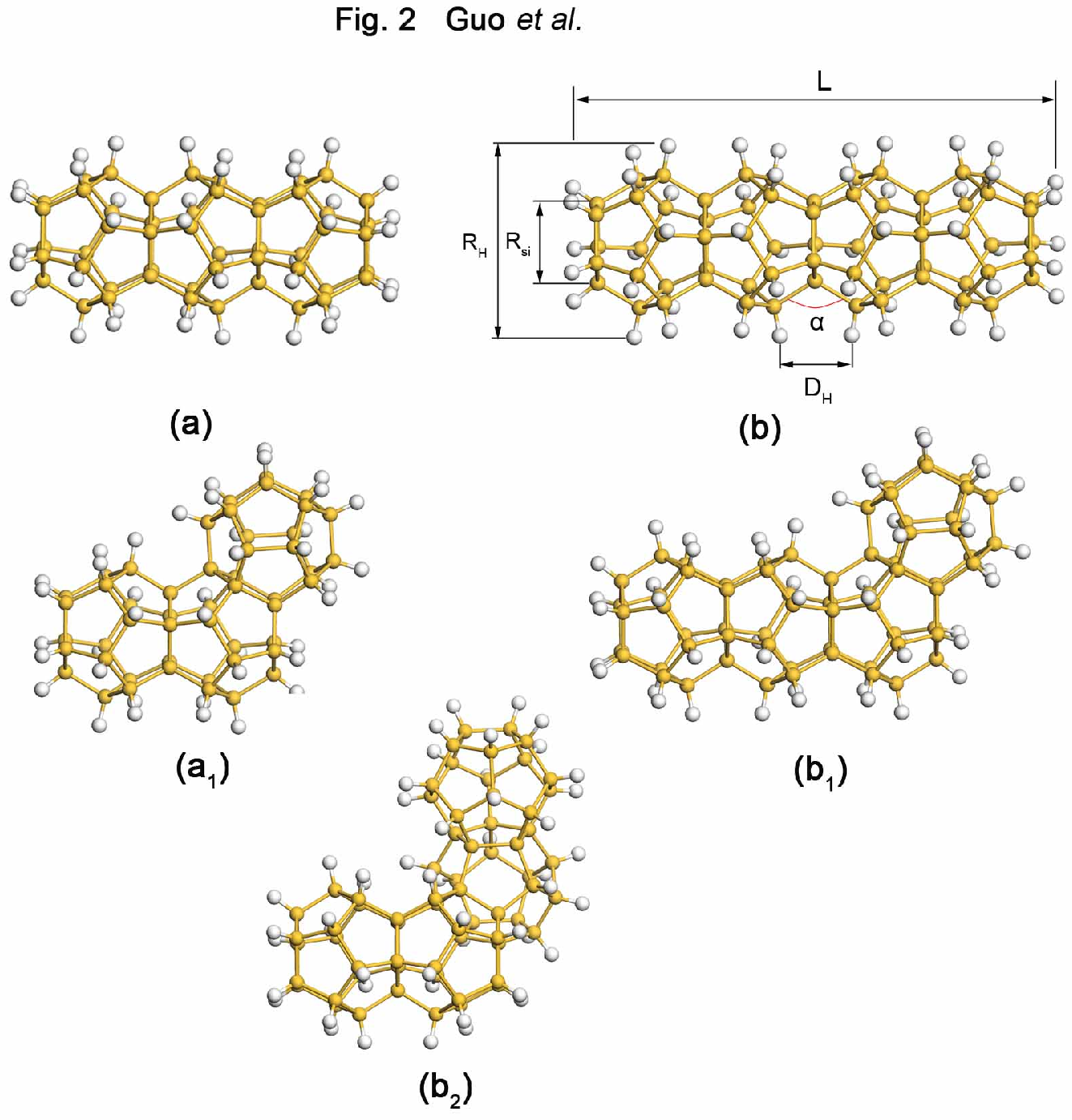}
\caption{\label{fig2}}
\end{figure}

\newpage
\begin{figure}
 \centering
\includegraphics[bb =0 0 700 800,width=1.40\textwidth]{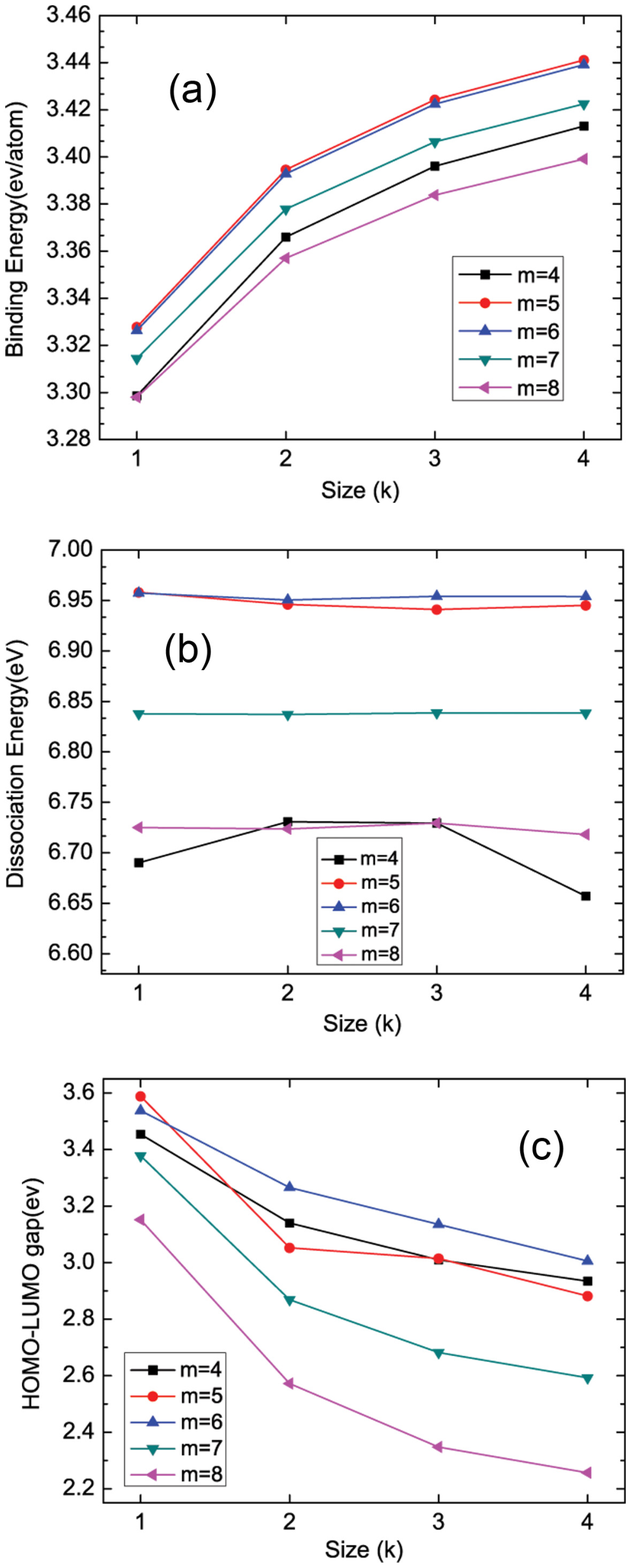}
\caption{\label{fig3}}
\end{figure}
\newpage
\begin{figure}
 \centering
\includegraphics[bb =80 0 650 450,width=1.70\textwidth]{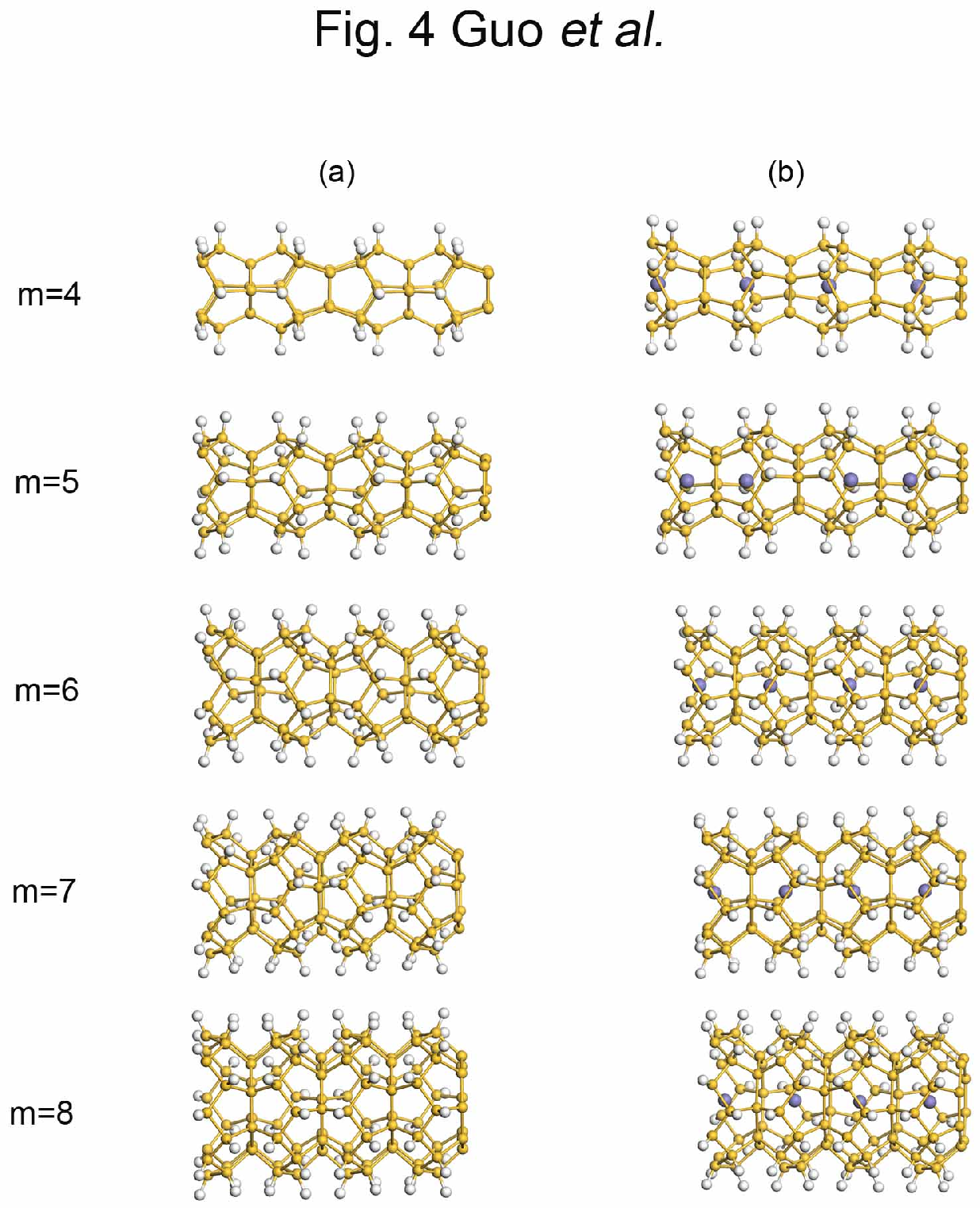}
\caption{\label{fig4}}
\end{figure}
\newpage
\begin{figure}
 \centering
\includegraphics[bb =100 100 900 900,width=1.20\textwidth]{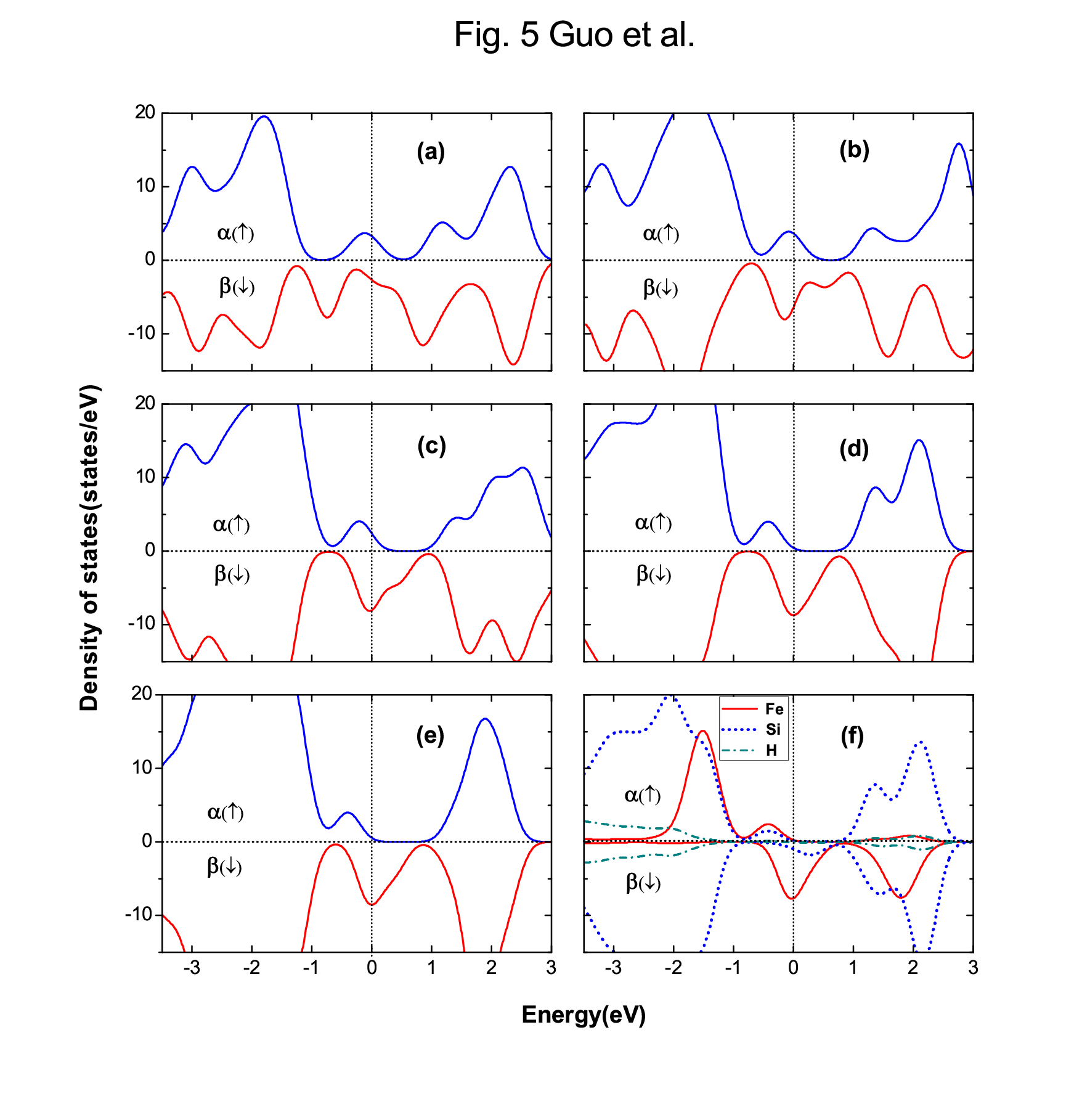}
\caption{\label{fig5}}
\end{figure}

\end{small}
\end{document}